\documentclass[final,5p,times,twocolumn]{elsarticle}

\usepackage{amssymb}
\usepackage{amsmath}
\usepackage{bm}
\usepackage{comment}
\usepackage[subrefformat=parens]{subcaption}
\usepackage{xcolor}
\usepackage{hyperref}
\usepackage{slashed}
\biboptions{sort&compress}
\hypersetup{
    colorlinks=true,       
    linkcolor=blue,          
    citecolor=blue,        
    urlcolor=blue,           
}

\journal{Physics Letters B}

\begin{document}

\begin{frontmatter}



\title{
  Ratio of baryon and electric-charge cumulants at second order
  with acceptance corrections
  \\ \vspace*{-16mm}\hspace{15.5cm} \small{\texttt{J-PARC-TH-0272}} \vspace*{14mm}
  }

\author[1,2]{Masakiyo Kitazawa\corref{cor1}}
\author[3]{ShinIchi Esumi}
\author[3]{Toshihiro Nonaka}

\cortext[cor1]{Corresponding author}

\address[1]{Department of Physics, Osaka University, Toyonaka, Osaka, 560-0043 Japan}
\address[2]{J-PARC Branch, KEK Theory Center, Institute of Particle and Nuclear Studies, KEK, 319-1106 Japan}
\address[3]{Tomonaga Center for the History of the Universe, University of Tsukuba, Tsukuba, Ibaraki, 305-8751, Japan}

\begin{abstract}
  We evaluate the ratio of baryon and electric-charge cumulants
  at second order from the recent experimental results
  at $\sqrt{s_{_{NN}}}=200$~GeV by the STAR Collaboration.
  The baryon number cumulant is reconstructed from the proton number
  distribution,
  and effects of the finite acceptance on the transverse momentum are
  corrected assuming the independent particle emission.
  We show that the obtained ratio has a dependence on the rapidity window.
  Comparison of the result with the hadron resonance gas model and
  lattice QCD numerical simulations suggests that if the fluctuations
  are generated from a thermal medium its temperature is 
  significantly lower than the chemical freezeout temperature.
\end{abstract}


\end{frontmatter}



\section{Introduction}
\label{sec:intro}

In relativistic heavy-ion collisions (HIC),
fluctuations are believed to be useful observables for investigating
phase transitions in the medium created by the
collisions~\cite{Asakawa:2015ybt,Bzdak:2019pkr,Bluhm:2020mpc}.
Fluctuations are characterized by cumulants~\cite{Asakawa:2015ybt},
which are known to show anomalous behaviors in a thermal medium near the
boundaries of QCD phase transitions especially as the order becomes
higher~\cite{Stephanov:1999zu,Asakawa:2000wh,Jeon:2000wg,Ejiri:2005wq,Stephanov:2008qz,Asakawa:2009aj,Friman:2011pf}.
In the HIC, fluctuations are measurable by the event-by-event analysis.
Various cumulants have been analyzed up to sixth order~\cite{STAR:2019ans,Adamczyk:2017wsl,STAR:2014egu,Adam:2020unf,STAR:2021iop,STAR:2021fge,Adamczewski-Musch:2020slf,ALICE:2019nbs},
and a suggestive non-monotonic behavior as functions of $\sqrt{s_{_{NN}}}$
has been reported~\cite{Adam:2020unf,STAR:2021iop}.
The experimental analyses will be refined further by the new data
from the RHIC-BES-II~\cite{BESIIwhitepaper} and 
HIC in future facilities~\cite{Galatyuk:2019lcf}.

Among fluctuation observables, those of conserved charges
have particularly useful properties.
First, the cumulants of conserved charges are calculable unambiguously
in thermal field theory, especially in lattice QCD numerical
simulations~\cite{Borsanyi:2014ewa,Ding:2015fca,DElia:2016jqh,Bazavov:2017dus,HotQCD:2017qwq,Borsanyi:2018grb,Bellwied:2019pxh,Bazavov:2020bjn,Bollweg:2021vqf}.
While the cumulants in a thermal system are extensive variables and
proportional to the spatial volume, the volume dependence can be eliminated
in their ratios~\cite{Ejiri:2005wq,Karsch:2010ck}.
The comparison between theoretical and experimental results
in terms of the ratios has been made in the
literature~\cite{Alba:2015iva,Chatterjee:2016mve,Bluhm:2018aei,Braun-Munzinger:2020jbk,Gupta:2022phu}.
Second, since the evolution of conserved charges is 
achieved only by the diffusion it is typically slow especially when
the spatial volume is taken to be large~\cite{Asakawa:2000wh,Jeon:2000wg,Asakawa:2019kek}.
In earlier studies~\cite{Asakawa:2000wh,Jeon:2000wg}, it has been
suggested to use this property of conserved-charge fluctuations
for investigating thermodynamics in the earlier stage of the HIC.
See Refs.~\cite{Kitazawa:2013bta,Kitazawa:2015ira} for an extension of
this idea for higher order cumulants.
Although fluctuations in the HIC are sometimes regarded to be
generated from a thermal medium around chemical freezeout, this assumption 
has to be checked carefully because of this property~\cite{Asakawa:2019kek}.

In the present Letter, among the conserved-charge cumulants
we focus on the ratio of the second-order cumulants of net-baryon number and
net-electric charge, $\langle N_{\rm B}^2 \rangle_{\rm c}$ and
$\langle N_{\rm Q}^2 \rangle_{\rm c}$.
An advantage to focus on this quantity
is that this would be the ratio between conserved-charge cumulants 
higher than the first order that can be analyzed most reliably in the HIC.
In the measurement of cumulants in the HIC, 
effects of the imperfect performance of detectors,
such as inefficiencies and finite acceptance,
have to be corrected~\cite{Kitazawa:2012at,Bzdak:2013pha,Nonaka:2017kko,Nonaka:2018mgw}.
The statistical and systematic uncertainties arising from this procedure,
however, grow rapidly as the order becomes higher.
Moreover, when comparing experimental results 
with theoretical ones assuming thermodynamics,
one has to consider modifications of fluctuations arising from 
experimental environment in the HIC, such as the 
volume fluctuations~\cite{Gorenstein:2011vq,Skokov:2012ds,Braun-Munzinger:2016yjz,Sugiura:2019toh,Adamczewski-Musch:2020slf}, 
global charge conservation~\cite{Sakaida:2014pya,Vovchenko:2020gne,Braun-Munzinger:2019yxj},
collision pileups~\cite{Garg:2017agr,Sombun:2017bxi,Nonaka:2020qcm}, 
dynamical evolution of fluctuations~\cite{Asakawa:2000wh,Jeon:2000wg,Kitazawa:2013bta,Kitazawa:2015ira}, and etc.
The modification of the cumulants due to these effects is more amplified
in a non-trivial way as the order becomes higher and
makes the meaningful comparison more difficult.
The use of the second-order cumulants enables a stable comparison
by suppressing these effects.

Another advantage to focus on this ratio is that, as we will see later,
in a thermal medium the ratio
$\langle N_{\rm B}^2 \rangle_{\rm c}/\langle N_{\rm Q}^2 \rangle_{\rm c}$
is a monotonically increasing function of temperature ($T$) and 
behaves almost linearly as a function of $T$
around the pseudo critical temperature $T_{\rm c}^*\simeq155$~MeV.
This linear $T$ dependence is suitable for studying the nature of
fluctuations in the HIC.

In the present study, we construct the values of 
$\langle N_{\rm B}^2 \rangle_{\rm c}$ and $\langle N_{\rm Q}^2 \rangle_{\rm c}$
and their ratio 
from the recent experimental data in Au+Au central collisions at $\sqrt{s_{_{NN}}}=200$~GeV by the STAR
Collaboration~\cite{STAR:2019ans,STAR:2021iop}.
In addition to the reconstruction of
baryon number cumulants~\cite{Kitazawa:2011wh,Kitazawa:2012at},
we perform the correction of the finite acceptance in the 
transverse momentum, $p_T$, space assuming the independent particle emission.
We show that the acceptance correction has a large effect
on the individual cumulants and the ratio, suggesting the
importance of the correction and the necessity to measure fluctuations
with wider acceptance.

We compare the value of 
$\langle N_{\rm B}^2 \rangle_{\rm c}/\langle N_{\rm Q}^2 \rangle_{\rm c}$
obtained in this way with
the ratio obtained in the hadron resonance gas (HRG) model
and the lattice QCD simulations.
Provided that the experimentally-observed fluctuations
are emitted from a thermal medium, 
the comparison shows the temperature $T\simeq134-138$~MeV,
which is significantly lower than the 
chemical freezeout temperature $T_{\rm chem}$.
Our result also shows that the ratio
$\langle N_{\rm B}^2 \rangle_{\rm c}/\langle N_{\rm Q}^2 \rangle_{\rm c}$
has a dependence on the rapidity window $\Delta y$ whereas
it is independent of $\Delta y$ for thermal fluctuations.
These results suggest the violation of the assumption
and motivate further investigations on the
nature of fluctuations in the HIC.

\section{Experimental data and their correction}
\label{sec:cumulants}

To obtain the cumulant ratio
$\langle N_{\rm B}^2 \rangle_{\rm c}/\langle N_{\rm Q}^2 \rangle_{\rm c}$
in the HIC,
we use experimental results by the STAR Collaboration in
Refs.~\cite{STAR:2019ans,STAR:2021iop}.
We construct $\langle N_{\rm B}^2 \rangle_{\rm c}$ from the data on the
proton number cumulants in Ref.~\cite{STAR:2021iop} according to the
procedure in Refs.~\cite{Kitazawa:2011wh,Kitazawa:2012at},
while we use the data in Ref.~\cite{STAR:2019ans}
for $\langle N_{\rm Q}^2 \rangle_{\rm c}$.
Effects of the detector's efficiencies are corrected in these results.
Throughout this study we concentrate on the result for the most central
($0-5\%$) Au+Au collisions at $\sqrt{s_{_{NN}}}=200$~GeV.
Effects of the violation of the boost invariance and the global charge
conservation are most suppressed and
the method in Refs.~\cite{Kitazawa:2011wh,Kitazawa:2012at}
is well justified at this $\sqrt{s_{_{NN}}}$.
Since the chemical potentials of conserved charges are small
at this $\sqrt{s_{_{NN}}}$, we neglect their
effects~\cite{HotQCD:2017qwq,Borsanyi:2018grb,Bazavov:2020bjn} in the following.
Since the effects of the QCD critical point~\cite{Stephanov:1999zu} would be
suppressed at this $\sqrt{s_{_{NN}}}$, the analysis is suitable for studying
non-critical behavior of fluctuations.

In Ref.~\cite{STAR:2021iop} and Ref.~\cite{STAR:2019ans},
particles are observed in {\it rapidity} and
{\it pseudo-rapidity} spaces, respectively.
To compare experimental results with theoretical ones,
the use of {\it space-time rapidity} is most desirable~\cite{Asakawa:2015ybt}.
Rapidity and pseudo-rapidity are used for its proxy,
while the former is better in the Bjorken picture~\cite{Ohnishi:2016bdf}.
This is the reason why we use the data on the proton number cumulant
in Ref.~\cite{STAR:2021iop}; for protons, rapidity and pseudo-rapidity
has a significant difference due to a heavy proton mass.
On the other hand, the difference is smaller 
in $\langle N_{\rm Q}^2 \rangle_{\rm c}$ since electric charges are
dominantly carried by pions whose mass is comparable with the mean $p_T$.
In the following, we thus regard the analysis in Ref.~\cite{STAR:2019ans}
as that in the rapidity space.

\begin{table}[tb]
  \centering
  \caption{
    Particle abundance in the $p_T$ acceptance $R_{p_T}$ obtained
    by the Blast-wave model in Ref.~\cite{STAR:2008med} for the
      most-central collisions at $\sqrt{s_{_{NN}}}=200$~GeV.
  }
  \label{table:RPT}
  \begin{tabular}{l|c}
    \hline
    particle species ($p_T$ range) & $R_{p_T}$ \\
    \hline
    pions ($0.4<p_T<1.6$~GeV)  & 0.44 \\
    kaons ($0.4<p_T<1.6$~GeV)  & 0.71 \\
    protons ($0.4<p_T<1.6$~GeV) & 0.71 \\
    $\pi$+$K$+$p$ ($0.4<p_T<1.6$~GeV) & 0.49 \\
    protons ($0.4<p_T<2.0$~GeV) & 0.82 \\
    \hline
  \end{tabular}
\end{table}

The measurements in Refs.~\cite{STAR:2019ans,STAR:2021iop} are performed
within a finite $p_T$-acceptance; 
$0.4<p_T<2.0$~GeV in Ref.~\cite{STAR:2021iop} and 
$0.4<p_T<1.6$~GeV in Ref.~\cite{STAR:2019ans}.
Due to the acceptance the particles in the final state are 
observed only with imperfect probabilities
\begin{align}
  R_{p_T} = \frac{{\rm (particle~number~in}~ p_T \mathrm{-acceptance)}}
  {\rm (total~particle~number)}.
  \label{eq:RPT}
\end{align}
Using the Blast wave model with the parameters 
for $\sqrt{s_{_{NN}}}=200$~GeV in Ref.~\cite{STAR:2008med},
the values of $R_{p_T}$ for individual particles are obtained
as shown in Table~\ref{table:RPT} for the $p_T$-acceptances
in Refs.~\cite{STAR:2019ans,STAR:2021iop}.
``$\pi$+$K$+$p$'' in the Table shows the weighted probability for the
charged particles with the particle abundances taken from Ref.~\cite{STAR:2008med}.

Since the measurement in the finite $p_T$-acceptance
modifies the particle-number distributions,
its effect on the cumulants has to be corrected
before comparing them with theoretical studies without such an acceptance cut.
In the present study, we perform this correction assuming that the individual
particles are emitted toward different $p_T$ with independent probabilities
according to a given $p_T$ distribution.
In this case, the correction can be carried out with the same procedure
as the efficiency correction assuming the binomial
distribution~\cite{Kitazawa:2012at,Asakawa:2015ybt}
with the probabilities $R_{p_T}$ in Table~\ref{table:RPT}\footnote{
  For $\langle N_{\rm Q}^2 \rangle_{\rm c}$, we carry out this procedure
  with the weighted probability, i.e. ``$\pi$+$K$+$p$'' in the Table.
  One may think that the correction procedure
  for multi-particle species~\cite{Nonaka:2017kko} with 
  individual probabilities for $\pi$, $K$, and $p$
  should be used for this correction.
  However, in Ref.~\cite{STAR:2019ans}
  the electric charge $N_{\rm Q}$ 
  is measured without particle identification.
  One thus cannot employ the method in Ref.~\cite{Nonaka:2017kko}
  without a detailed knowledge on the detector's response.
  In any case, as discussed in Ref.~\cite{Nonaka:2017kko}
  the systematic deviation due to the use of the weighted $R_{p_T}$
  is small at the second order and thus our results would be less
  affected,
  while the effect is amplified for higher order cumulants.
}.

We note that another way to compensate the effect of the $p_T$-acceptance 
in the comparison between experimental and theoretical analyses is
to perform the correction in theoretical calculations, if possible.
Such comparisons have been made in the studies employing the HRG
model~\cite{Alba:2014eba,Alba:2015iva,Bluhm:2018aei,Bellwied:2019pxh,Vovchenko:2022szk}.
However, in general it is not possible to perform such a correction
in theoretical calculations.
The correction of the experimental data enables direct comparisons 
even for such cases in terms of the true values of cumulants.

\begin{figure*}[tbp]
  \centering
  \includegraphics[width=0.49\textwidth]{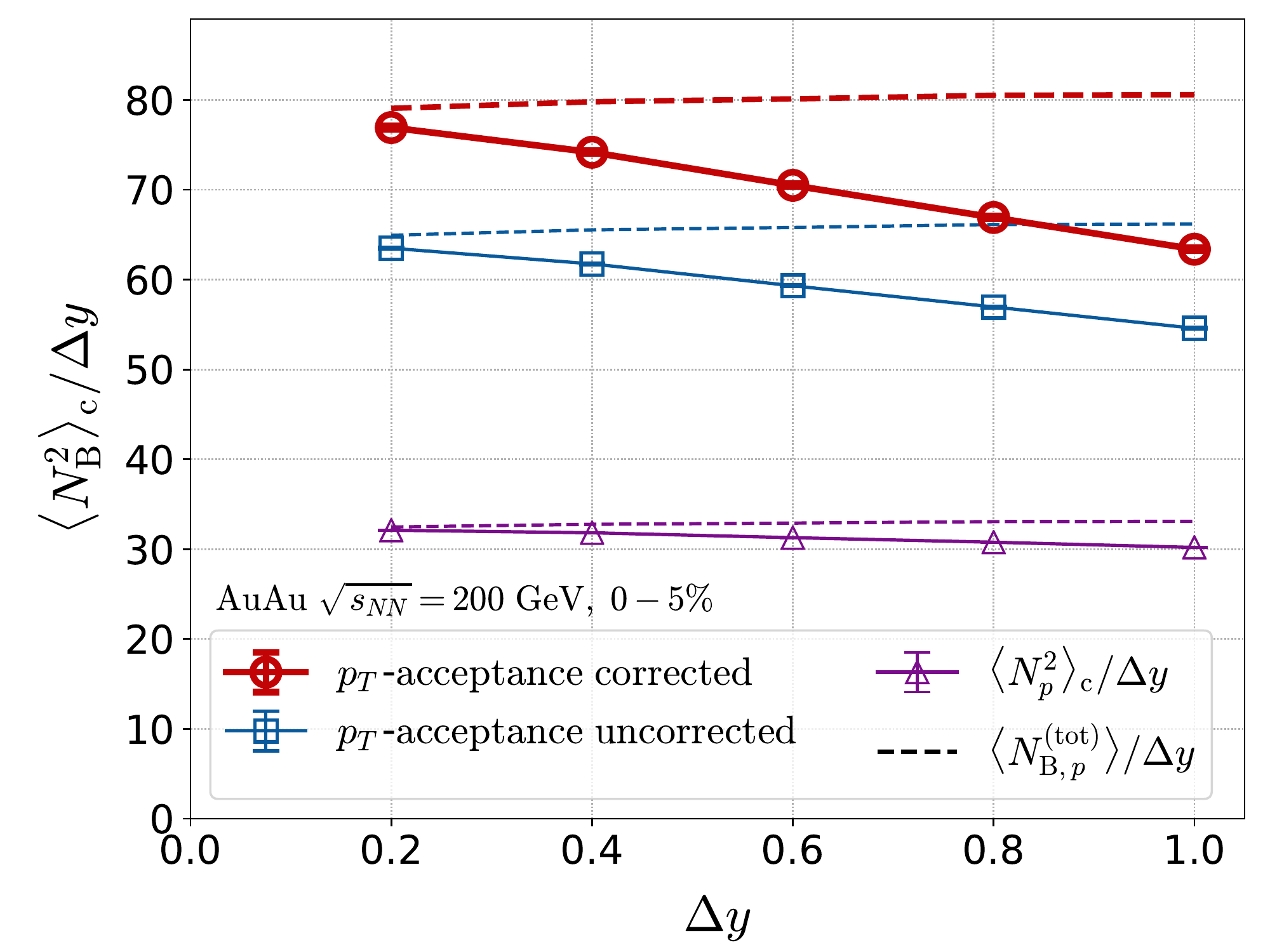}
  \includegraphics[width=0.49\textwidth]{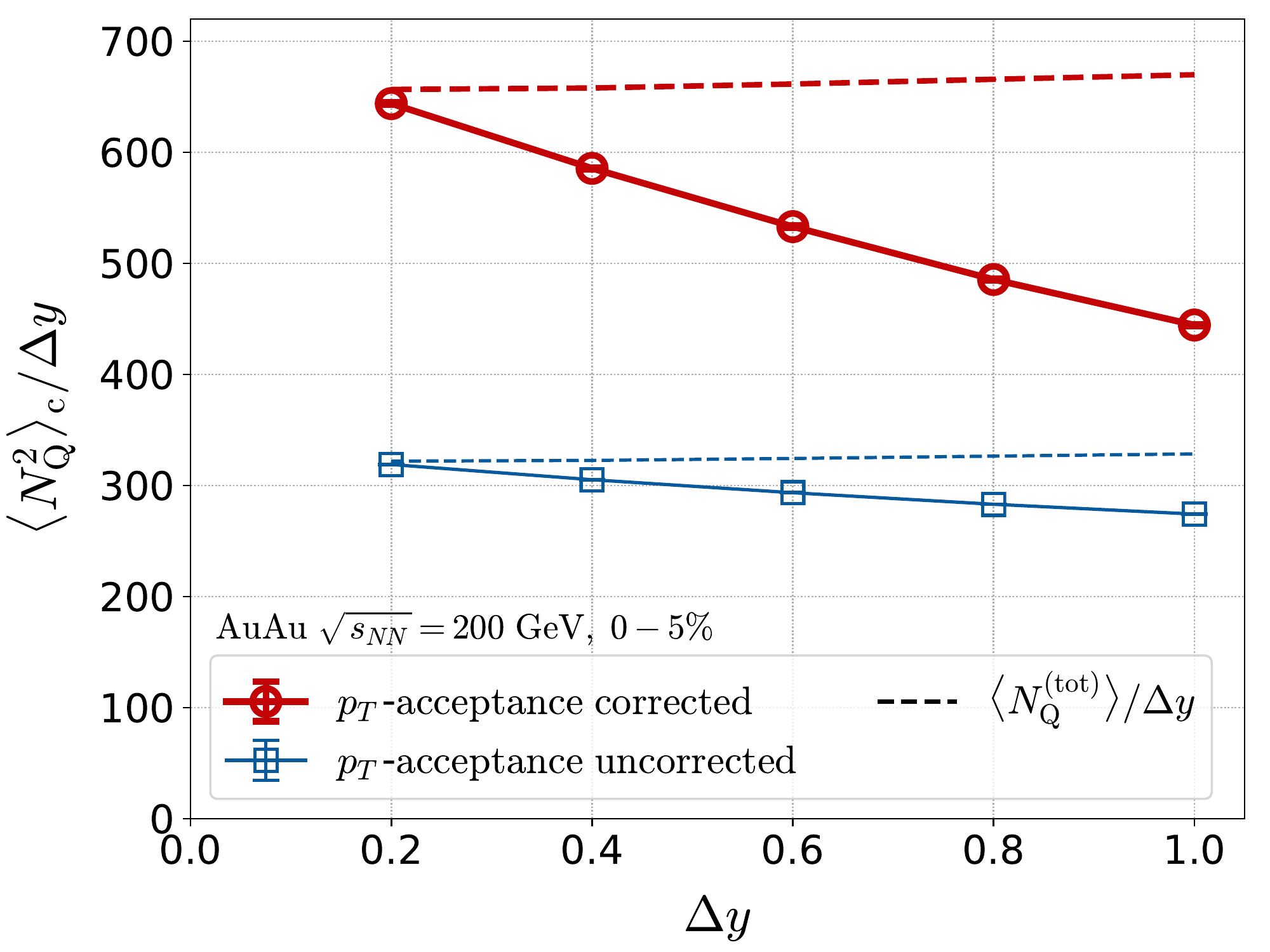}
  \caption{
    Second-order cumulants $\langle N_{\rm B}^2 \rangle_{\rm c}$,
    $\langle N_p^2 \rangle_{\rm c}$, and $\langle N_{\rm Q}^2 \rangle_{\rm c}$
    divided by $\Delta y$ obtained from the experimental results
    in Refs.~\cite{STAR:2019ans,STAR:2021iop} at $\sqrt{s_{_{NN}}}=200$~GeV.
    The circle and square symbols show the results with and without the
    $p_T$-acceptance correction, respectively.
    The triangles in the right panel show $\langle N_p^2 \rangle_{\rm c}/\Delta y$.
    The error bars show statistical errors.
    The dashed lines near the symbols are the corresponding
    total particle numbers.
  }
  \label{fig:chiBQ}
\end{figure*}

\section{Cumulant ratio}
\label{sec:result}

In Fig.~\ref{fig:chiBQ}, we show the second-order cumulants
$\langle N_{\rm B}^2 \rangle_{\rm c}$, $\langle N_{\rm Q}^2 \rangle_{\rm c}$,
as well as that of the proton number $\langle N_p^2 \rangle_{\rm c}$, 
divided by the rapidity window $\Delta y$ as functions of $\Delta y$.
These quantities are constant if they are generated from a thermal system
having a boost invariance~\cite{Asakawa:2015ybt}.
In the left panel, the triangles show $\langle N_p^2 \rangle_{\rm c}/\Delta y$
in Ref.~\cite{STAR:2021iop}.
The dashed line near the data shows the total particle number
$\langle N_p^{\rm (total)} \rangle/\Delta y$.
The squares in the same panel show 
$\langle N_{\rm B}^2 \rangle_{\rm c}/\Delta y$ obtained with 
the procedure in Refs.~\cite{Kitazawa:2011wh,Kitazawa:2012at},
while the circles show $\langle N_{\rm B}^2 \rangle_{\rm c}/\Delta y$
for which the $p_T$-acceptance correction is performed.
The error bars show the statistical errors, 
which are negligibly small in these results.
The dashed lines near these results are the total baryon number 
$\langle N_{\rm B}^{\rm (total)} \rangle/\Delta y= 2\langle N_p^{\rm (total)} \rangle/\Delta y$.
One sees that the deviation of $\langle N_{\rm B}^2 \rangle_{\rm c}$
from $\langle N_{\rm B}^{\rm (total)} \rangle$
at large $\Delta y$ is pronounced by the corrections.
This result is reasonable since the incomplete measurement tends to make the
distribution close to the Skellam distribution in which
$\langle N_{\rm B}^2 \rangle_{\rm c}=\langle N_{\rm B}^{\rm (total)} \rangle$~\cite{Asakawa:2015ybt}.

Shown in the right panel of Fig.~\ref{fig:chiBQ} are 
$\langle N_{\rm Q}^2 \rangle_{\rm c}/\Delta y$ 
with (circles) and without (squares) the $p_T$-acceptance correction\footnote{
  The total electric charge required for the $p_T$-acceptance correction
  is provided by the private communication with Arghya Chatterjee.
}.
The meaning of the dashed lines is the same as the left panel.
The panel shows that the effect of the $p_T$-acceptance correction is
more significant than $\langle N_{\rm B}^2 \rangle_{\rm c}$ 
because of the smaller $R_{p_T}$ for the electric charge.

\begin{figure*}[tbp]
  \centering
  \includegraphics[width=0.8\textwidth]{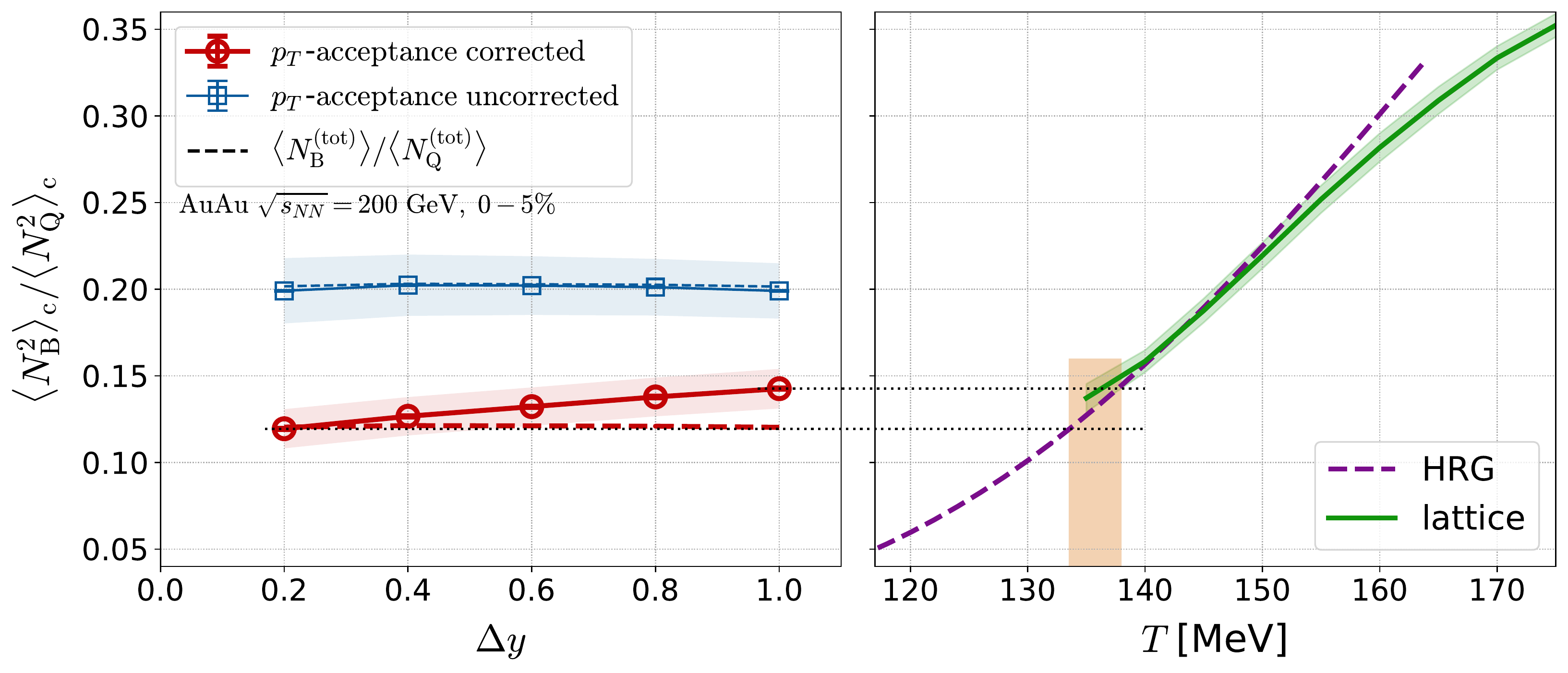}
  \caption{
    Left panel shows the ratio 
    $\langle \delta N_{\rm B}^2 \rangle / \langle \delta N_{\rm Q}^2 \rangle$
    obtained from the HIC as a function of the rapidity window $\Delta y$.
    The right panel shows the ratio obtained in the lattice QCD
    simulations~\cite{Bollweg:2021vqf} and the HRG model.
    The dotted horizontal lines and shaded area show a comparison
    of the two results.
  }
  \label{fig:comp}
\end{figure*}

In the left panel of Fig.~\ref{fig:comp}, we show the
$p_T$-acceptance-corrected result of 
$\langle N_{\rm B}^2 \rangle_{\rm c}/\langle N_{\rm Q}^2 \rangle_{\rm c}$
by the circles.
The dashed lines show
$\langle N_{\rm B}^{\rm (total)} \rangle/\langle N_{\rm Q}^{\rm (total)} \rangle$.
The shaded band represents the systematic errors that account for 
the propagation from that of $\langle N_{\rm B}^2 \rangle_{\rm c}$
in Refs.~\cite{STAR:2021iop}; 
we, however, note that this error band should be regarded
only as a guide since the estimate of the systematic errors of 
$\langle N_{\rm B}^2 \rangle_{\rm c}/\langle N_{\rm Q}^2 \rangle_{\rm c}$
needs a detailed knowledge on the experimental analyses.
In the panel, the result without the $p_T$-acceptance correction is also
shown by the squares as a reference.
One sees that the correction strongly modifies the ratio.

If fluctuations are emitted from a thermal system having a boost invariance,
the ratio $\langle N_{\rm B}^2 \rangle_{\rm c}/\langle N_{\rm Q}^2 \rangle_{\rm c}$
is a constant as a function of $\Delta y$~\cite{Asakawa:2015ybt}.
While $\langle N_{\rm B}^2 \rangle_{\rm c}/\Delta y$ and 
$\langle N_{\rm Q}^2 \rangle_{\rm c}/\Delta y$ become decreasing functions
when the effects of the global charge conservation are taken into account~\cite{Sakaida:2014pya,Vovchenko:2020gne},
these $\Delta y$ dependence cancels out in the ratio
$\langle N_{\rm B}^2 \rangle_{\rm c}/\langle N_{\rm Q}^2 \rangle_{\rm c}$
for the thermal case~\cite{Vovchenko:2020gne}.
On the other hand, from the left panel of Fig.~\ref{fig:comp} one sees that
the acceptance-corrected ratio 
has a clear increasing trend as a function of $\Delta y$.
This result shows that the fluctuations in the HIC
are not emitted from a purely thermal system
including the effect of global conservation.

\section{Comparison with HRG model and lattice results}
\label{sec:HRG}

Assuming that the fluctuations observed in the HIC
are those of a thermal system, one can estimate the temperature of
the system by comparing the ratio of cumulants with the results
obtained in lattice QCD simulations.
Even when the fluctuations are not thermal, such a comparison is 
useful for investigating the nature of the fluctuations.

To perform the comparison using the ratio
$\langle N_{\rm B}^2 \rangle_{\rm c}/\langle N_{\rm Q}^2 \rangle_{\rm c}$,
in the right panel of Fig.~\ref{fig:comp} we show the $T$ dependence
of $\langle N_{\rm B}^2 \rangle_{\rm c}/\langle N_{\rm Q}^2 \rangle_{\rm c}$
obtained from a lattice QCD simulation~\cite{Bollweg:2021vqf}
by the solid line with an error band.
Finite-volume effects of $\langle N_{\rm Q}^2 \rangle_{\rm c}$ 
are corrected according to Ref.~\cite{Bollweg:2021vqf}.
The range of the vertical axis is the same as the left panel.
The lattice results on thermodynamics are known to be well reproduced
by the HRG model at low $T$.
In the panel, the ratio $\langle N_{\rm B}^2 \rangle_{\rm c}/\langle N_{\rm Q}^2 \rangle_{\rm c}$
obtained in the HRG model is shown by the dashed line,
where we use the set of hadrons in ``QMHRG2020''~\cite{Bollweg:2021vqf}
for the HRG model.
The figure shows that the lattice result agrees well with the HRG model
for $T\lesssim 145$~MeV, which suggests the validity of the latter
in this range of $T$.
From the panel one also finds that 
$\langle N_{\rm B}^2 \rangle_{\rm c}/\langle N_{\rm Q}^2 \rangle_{\rm c}$
behaves almost linearly 
as a function of $T$ in the range of $T$ shown in the panel.
As discussed in Sec.~\ref{sec:intro},
this is an attractive feature of this ratio.

To compare the results in the left and right panels, 
in Fig.~\ref{fig:comp} we show the dotted horizontal lines 
at the values of
$\langle N_{\rm B}^2 \rangle_{\rm c}/\langle N_{\rm Q}^2 \rangle_{\rm c}$
obtained from the HIC at $\Delta y=0.2$ and $1.0$.
By comparing these values with the ratio in the HRG model,
one finds that the temperature extracted from the na\"ive comparison gives
$T\simeq134-138$~MeV depending on $\Delta y$ as shown by the shaded box
in the right panel.
We note that this temperature is significantly smaller than the chemical
freezeout temperature $T_{\rm chem}\simeq156$~MeV
for $\sqrt{s_{_{NN}}}=200$~GeV~\cite{STAR:2008med}.
The value of $\langle N_{\rm B}^2 \rangle_{\rm c}/\langle N_{\rm Q}^2 \rangle_{\rm c}$
in the HIC itself is about twice smaller than
the value in the HRG model at $T=T_{\rm chem}$.

\section{Discussions}
\label{sec:discussion}

In the present study, we have investigated the cumulant ratio
$\langle N_{\rm B}^2 \rangle_{\rm c}/\langle N_{\rm Q}^2 \rangle_{\rm c}$
observed in the HIC.
Because this ratio consists only of the second-order conserved-charge
cumulants, various uncertainties in the experimental measurement
that are amplified for higher order cumulants are suppressed in 
its analysis.
The ratio in the HIC at $\sqrt{s_{_{NN}}}=200$~GeV is
estimated from the experimental results
by the STAR Collaboration~\cite{STAR:2019ans,STAR:2021iop}.
In addition to the reconstruction of the baryon number cumulant
from those of protons, effects of the $p_T$-acceptance are corrected 
assuming the independent particle emission.
Our result shows that this correction strongly modifies the resulting
values of the cumulants and their ratio and thus is crucial.
Since the effect of the correction becomes more significant
for higher order cumulants~\cite{Nonaka:2017kko},
this result also shows the importance of the correction 
in their analysis for the search for the QCD critical point.

The na\"ive comparison of the obtained ratio with the HRG model
suggests the temperature $T\simeq134-138$~MeV, which is significantly
lower than $T_{\rm chem}$.
By taking this result seriously, it is suggested that 
the fluctuation observables in the HIC are generated in the hadronic
phase later than the chemical freeze out
in contrast to the earlier suggestions~\cite{Asakawa:2000wh,Jeon:2000wg}.

However, we emphasize that this comparison is made assuming that 
the fluctuations in the HIC are thermal.
On the other hand, existence of the $\Delta y$ dependence of
$\langle N_{\rm B}^2\rangle_{\rm c}/\langle N_{\rm Q}^2\rangle_{\rm c}$ 
shows the violation of this assumption in the HIC.
Therefore, to understand the experimental result correctly
one needs further investigations on the nature of fluctuations especially 
taking their dynamics into account~\cite{Asakawa:2000wh,Jeon:2000wg,Kitazawa:2013bta,Kitazawa:2015ira}.
The modifications of the cumulants due to the 
use of (pseudo-)rapidity in place of space-time rapidity~\cite{Ohnishi:2016bdf}
and the resonance decays after the chemical freezeout are 
other important effects to be considered 
since they tend to make $\langle N_{\rm Q}^2\rangle_{\rm c}$ larger 
and suppress $\langle N_{\rm B}^2\rangle_{\rm c}/\langle N_{\rm Q}^2\rangle_{\rm c}$.
Because these effects are suppressed by extending $\Delta y$~\cite{Asakawa:2015ybt},
the measurement of the fluctuations with larger $\Delta y$ is an important
experimental subject for resolving these issues.

Finally, we remark that the value of 
$\langle N_{\rm B}^2\rangle_{\rm c}/\langle N_{\rm Q}^2\rangle_{\rm c}$
obtained in the present study from Refs.~\cite{STAR:2019ans,STAR:2021iop}
would have a deviation from the true value.
As discussed in Sec.~\ref{sec:cumulants}, while
$\langle N_{\rm Q}^2\rangle_{\rm c}$ is measured in the pseudo-rapidity space 
in Ref.~\cite{STAR:2019ans}, the measurement in the rapidity space is
more desirable.
Although we have performed the $p_T$-acceptance correction
assuming the independent particle emission in the $p_T$ space,
there is no a priori justification of this assumption in the HIC.
In addition to theoretical studies on this point, 
the measurement of fluctuations with wider acceptance is an important
experimental subject 
for reducing uncertainties from the violation of this assumption.
As emphasized already, such measurements are more important for 
the analyses of higher order cumulants.
It thus is quite interesting to realize such experiments
at the future experiments at FAIR, NICA and J-PARC-HI~\cite{Galatyuk:2019lcf}.

\section*{Acknowledgments}

The authors thank Arghya Chatterjee for providing unpublished
data in Ref.~\cite{STAR:2019ans}.
They also thank Krzysztof Redlich and Volodymyr Vovchenko for
valuable comments.
This work was supported by JSPS KAKENHI 
(Grants No.~JP19H05598, JP20H01903, JP22K03619).

\bibliographystyle{elsarticle-num} 
\bibliography{refs.bib}

\begin{thebibliography}{10}
\expandafter\ifx\csname url\endcsname\relax
  \def\url#1{\texttt{#1}}\fi
\expandafter\ifx\csname urlprefix\endcsname\relax\def\urlprefix{URL }\fi
\expandafter\ifx\csname href\endcsname\relax
  \def\href#1#2{#2} \def\path#1{#1}\fi

\bibitem{Asakawa:2015ybt}
M.~Asakawa, M.~Kitazawa, {Fluctuations of conserved charges in relativistic
  heavy ion collisions: An introduction}, Prog. Part. Nucl. Phys. 90 (2016)
  299--342.
\newblock \href {http://arxiv.org/abs/1512.05038} {\path{arXiv:1512.05038}},
  \href {https://doi.org/10.1016/j.ppnp.2016.04.002}
  {\path{doi:10.1016/j.ppnp.2016.04.002}}.

\bibitem{Bzdak:2019pkr}
A.~Bzdak, S.~Esumi, V.~Koch, J.~Liao, M.~Stephanov, N.~Xu, {Mapping the Phases
  of Quantum Chromodynamics with Beam Energy Scan}, Phys. Rept. 853 (2020)
  1--87.
\newblock \href {http://arxiv.org/abs/1906.00936} {\path{arXiv:1906.00936}},
  \href {https://doi.org/10.1016/j.physrep.2020.01.005}
  {\path{doi:10.1016/j.physrep.2020.01.005}}.

\bibitem{Bluhm:2020mpc}
M.~Bluhm, et~al., {Dynamics of critical fluctuations: Theory \textendash{}
  phenomenology \textendash{} heavy-ion collisions}, Nucl. Phys. A 1003 (2020)
  122016.
\newblock \href {http://arxiv.org/abs/2001.08831} {\path{arXiv:2001.08831}},
  \href {https://doi.org/10.1016/j.nuclphysa.2020.122016}
  {\path{doi:10.1016/j.nuclphysa.2020.122016}}.

\bibitem{Stephanov:1999zu}
M.~A. Stephanov, K.~Rajagopal, E.~V. Shuryak, {Event-by-event fluctuations in
  heavy ion collisions and the QCD critical point}, Phys. Rev. D 60 (1999)
  114028.
\newblock \href {http://arxiv.org/abs/hep-ph/9903292}
  {\path{arXiv:hep-ph/9903292}}, \href
  {https://doi.org/10.1103/PhysRevD.60.114028}
  {\path{doi:10.1103/PhysRevD.60.114028}}.

\bibitem{Asakawa:2000wh}
M.~Asakawa, U.~W. Heinz, B.~Muller, {Fluctuation probes of quark
  deconfinement}, Phys. Rev. Lett. 85 (2000) 2072--2075.
\newblock \href {http://arxiv.org/abs/hep-ph/0003169}
  {\path{arXiv:hep-ph/0003169}}, \href
  {https://doi.org/10.1103/PhysRevLett.85.2072}
  {\path{doi:10.1103/PhysRevLett.85.2072}}.

\bibitem{Jeon:2000wg}
S.~Jeon, V.~Koch, {Charged particle ratio fluctuation as a signal for QGP},
  Phys. Rev. Lett. 85 (2000) 2076--2079.
\newblock \href {http://arxiv.org/abs/hep-ph/0003168}
  {\path{arXiv:hep-ph/0003168}}, \href
  {https://doi.org/10.1103/PhysRevLett.85.2076}
  {\path{doi:10.1103/PhysRevLett.85.2076}}.

\bibitem{Ejiri:2005wq}
S.~Ejiri, F.~Karsch, K.~Redlich, {Hadronic fluctuations at the QCD phase
  transition}, Phys. Lett. B 633 (2006) 275--282.
\newblock \href {http://arxiv.org/abs/hep-ph/0509051}
  {\path{arXiv:hep-ph/0509051}}, \href
  {https://doi.org/10.1016/j.physletb.2005.11.083}
  {\path{doi:10.1016/j.physletb.2005.11.083}}.

\bibitem{Stephanov:2008qz}
M.~A. Stephanov, {Non-Gaussian fluctuations near the QCD critical point}, Phys.
  Rev. Lett. 102 (2009) 032301.
\newblock \href {http://arxiv.org/abs/0809.3450} {\path{arXiv:0809.3450}},
  \href {https://doi.org/10.1103/PhysRevLett.102.032301}
  {\path{doi:10.1103/PhysRevLett.102.032301}}.

\bibitem{Asakawa:2009aj}
M.~Asakawa, S.~Ejiri, M.~Kitazawa, {Third moments of conserved charges as
  probes of QCD phase structure}, Phys. Rev. Lett. 103 (2009) 262301.
\newblock \href {http://arxiv.org/abs/0904.2089} {\path{arXiv:0904.2089}},
  \href {https://doi.org/10.1103/PhysRevLett.103.262301}
  {\path{doi:10.1103/PhysRevLett.103.262301}}.

\bibitem{Friman:2011pf}
B.~Friman, F.~Karsch, K.~Redlich, V.~Skokov, {Fluctuations as probe of the QCD
  phase transition and freeze-out in heavy ion collisions at LHC and RHIC},
  Eur. Phys. J. C 71 (2011) 1694.
\newblock \href {http://arxiv.org/abs/1103.3511} {\path{arXiv:1103.3511}},
  \href {https://doi.org/10.1140/epjc/s10052-011-1694-2}
  {\path{doi:10.1140/epjc/s10052-011-1694-2}}.

\bibitem{STAR:2019ans}
J.~Adam, et~al., {Collision-energy dependence of second-order off-diagonal and
  diagonal cumulants of net-charge, net-proton, and net-kaon multiplicity
  distributions in Au + Au collisions}, Phys. Rev. C 100~(1) (2019) 014902,
  [Erratum: Phys.Rev.C 105, 029901 (2022)].
\newblock \href {http://arxiv.org/abs/1903.05370} {\path{arXiv:1903.05370}},
  \href {https://doi.org/10.1103/PhysRevC.100.014902}
  {\path{doi:10.1103/PhysRevC.100.014902}}.

\bibitem{Adamczyk:2017wsl}
L.~Adamczyk, et~al., {Collision Energy Dependence of Moments of Net-Kaon
  Multiplicity Distributions at RHIC} (2017).
\newblock \href {http://arxiv.org/abs/1709.00773} {\path{arXiv:1709.00773}}.

\bibitem{STAR:2014egu}
L.~Adamczyk, et~al., {Beam energy dependence of moments of the net-charge
  multiplicity distributions in Au+Au collisions at RHIC}, Phys. Rev. Lett. 113
  (2014) 092301.
\newblock \href {http://arxiv.org/abs/1402.1558} {\path{arXiv:1402.1558}},
  \href {https://doi.org/10.1103/PhysRevLett.113.092301}
  {\path{doi:10.1103/PhysRevLett.113.092301}}.

\bibitem{Adam:2020unf}
J.~Adam, et~al., {Net-proton number fluctuations and the Quantum Chromodynamics
  critical point} (2020).
\newblock \href {http://arxiv.org/abs/2001.02852} {\path{arXiv:2001.02852}}.

\bibitem{STAR:2021iop}
M.~Abdallah, et~al., {Cumulants and correlation functions of net-proton,
  proton, and antiproton multiplicity distributions in Au+Au collisions at
  energies available at the BNL Relativistic Heavy Ion Collider}, Phys. Rev. C
  104~(2) (2021) 024902.
\newblock \href {http://arxiv.org/abs/2101.12413} {\path{arXiv:2101.12413}},
  \href {https://doi.org/10.1103/PhysRevC.104.024902}
  {\path{doi:10.1103/PhysRevC.104.024902}}.

\bibitem{STAR:2021fge}
M.~S. Abdallah, et~al., {Measurements of Proton High Order Cumulants in 3 GeV
  Au+Au Collisions and Implications for the QCD Critical Point} (11 2021).
\newblock \href {http://arxiv.org/abs/2112.00240} {\path{arXiv:2112.00240}}.

\bibitem{Adamczewski-Musch:2020slf}
J.~Adamczewski-Musch, et~al., {Proton number fluctuations in $\sqrt{s_{NN}}$ =
  2.4 GeV Au+Au collisions studied with HADES} (2020).
\newblock \href {http://arxiv.org/abs/2002.08701} {\path{arXiv:2002.08701}}.

\bibitem{ALICE:2019nbs}
S.~Acharya, et~al., {Global baryon number conservation encoded in net-proton
  fluctuations measured in Pb-Pb collisions at $\sqrt{s_{\rm NN}}$ = 2.76 TeV},
  Phys. Lett. B 807 (2020) 135564.
\newblock \href {http://arxiv.org/abs/1910.14396} {\path{arXiv:1910.14396}},
  \href {https://doi.org/10.1016/j.physletb.2020.135564}
  {\path{doi:10.1016/j.physletb.2020.135564}}.

\bibitem{BESIIwhitepaper}
\href{https://drupal.star.bnl.gov/STAR/system/files/BES_WPII_ver6.9_Cover.pdf}{A
  {STAR} white paper summarizing the current understanding and describing
  future plans} (2014).
\newline\urlprefix\url{https://drupal.star.bnl.gov/STAR/system/files/BES_WPII_ver6.9_Cover.pdf}

\bibitem{Galatyuk:2019lcf}
T.~Galatyuk, {Future facilities for high $\mu_B$ physics}, Nucl. Phys. A 982
  (2019) 163--169.
\newblock \href {https://doi.org/10.1016/j.nuclphysa.2018.11.025}
  {\path{doi:10.1016/j.nuclphysa.2018.11.025}}.

\bibitem{Borsanyi:2014ewa}
S.~Borsanyi, Z.~Fodor, S.~D. Katz, S.~Krieg, C.~Ratti, K.~K. Szabo, {Freeze-out
  parameters from electric charge and baryon number fluctuations: is there
  consistency?}, Phys. Rev. Lett. 113 (2014) 052301.
\newblock \href {http://arxiv.org/abs/1403.4576} {\path{arXiv:1403.4576}},
  \href {https://doi.org/10.1103/PhysRevLett.113.052301}
  {\path{doi:10.1103/PhysRevLett.113.052301}}.

\bibitem{Ding:2015fca}
H.~T. Ding, S.~Mukherjee, H.~Ohno, P.~Petreczky, H.~P. Schadler, {Diagonal and
  off-diagonal quark number susceptibilities at high temperatures}, Phys. Rev.
  D 92~(7) (2015) 074043.
\newblock \href {http://arxiv.org/abs/1507.06637} {\path{arXiv:1507.06637}},
  \href {https://doi.org/10.1103/PhysRevD.92.074043}
  {\path{doi:10.1103/PhysRevD.92.074043}}.

\bibitem{DElia:2016jqh}
M.~D'Elia, G.~Gagliardi, F.~Sanfilippo, {Higher order quark number fluctuations
  via imaginary chemical potentials in $N_f=2+1$ QCD}, Phys. Rev. D 95~(9)
  (2017) 094503.
\newblock \href {http://arxiv.org/abs/1611.08285} {\path{arXiv:1611.08285}},
  \href {https://doi.org/10.1103/PhysRevD.95.094503}
  {\path{doi:10.1103/PhysRevD.95.094503}}.

\bibitem{Bazavov:2017dus}
A.~Bazavov, et~al., {The QCD Equation of State to $\mathcal{O}(\mu_B^6)$ from
  Lattice QCD}, Phys. Rev. D 95~(5) (2017) 054504.
\newblock \href {http://arxiv.org/abs/1701.04325} {\path{arXiv:1701.04325}},
  \href {https://doi.org/10.1103/PhysRevD.95.054504}
  {\path{doi:10.1103/PhysRevD.95.054504}}.

\bibitem{HotQCD:2017qwq}
A.~Bazavov, et~al., {Skewness and kurtosis of net baryon-number distributions
  at small values of the baryon chemical potential}, Phys. Rev. D 96~(7) (2017)
  074510.
\newblock \href {http://arxiv.org/abs/1708.04897} {\path{arXiv:1708.04897}},
  \href {https://doi.org/10.1103/PhysRevD.96.074510}
  {\path{doi:10.1103/PhysRevD.96.074510}}.

\bibitem{Borsanyi:2018grb}
S.~Borsanyi, Z.~Fodor, J.~N. Guenther, S.~K. Katz, K.~K. Szabo, A.~Pasztor,
  I.~Portillo, C.~Ratti, {Higher order fluctuations and correlations of
  conserved charges from lattice QCD}, JHEP 10 (2018) 205.
\newblock \href {http://arxiv.org/abs/1805.04445} {\path{arXiv:1805.04445}},
  \href {https://doi.org/10.1007/JHEP10(2018)205}
  {\path{doi:10.1007/JHEP10(2018)205}}.

\bibitem{Bellwied:2019pxh}
R.~Bellwied, S.~Borsanyi, Z.~Fodor, J.~N. Guenther, J.~Noronha-Hostler,
  P.~Parotto, A.~Pasztor, C.~Ratti, J.~M. Stafford, {Off-diagonal correlators
  of conserved charges from lattice QCD and how to relate them to experiment},
  Phys. Rev. D 101~(3) (2020) 034506.
\newblock \href {http://arxiv.org/abs/1910.14592} {\path{arXiv:1910.14592}},
  \href {https://doi.org/10.1103/PhysRevD.101.034506}
  {\path{doi:10.1103/PhysRevD.101.034506}}.

\bibitem{Bazavov:2020bjn}
A.~Bazavov, et~al., {Skewness, kurtosis, and the fifth and sixth order
  cumulants of net baryon-number distributions from lattice QCD confront
  high-statistics STAR data}, Phys. Rev. D 101~(7) (2020) 074502.
\newblock \href {http://arxiv.org/abs/2001.08530} {\path{arXiv:2001.08530}},
  \href {https://doi.org/10.1103/PhysRevD.101.074502}
  {\path{doi:10.1103/PhysRevD.101.074502}}.

\bibitem{Bollweg:2021vqf}
D.~Bollweg, J.~Goswami, O.~Kaczmarek, F.~Karsch, S.~Mukherjee, P.~Petreczky,
  C.~Schmidt, P.~Scior, {Second order cumulants of conserved charge
  fluctuations revisited: Vanishing chemical potentials}, Phys. Rev. D 104~(7)
  (2021).
\newblock \href {http://arxiv.org/abs/2107.10011} {\path{arXiv:2107.10011}},
  \href {https://doi.org/10.1103/PhysRevD.104.074512}
  {\path{doi:10.1103/PhysRevD.104.074512}}.

\bibitem{Karsch:2010ck}
F.~Karsch, K.~Redlich, {Probing freeze-out conditions in heavy ion collisions
  with moments of charge fluctuations}, Phys. Lett. B 695 (2011) 136--142.
\newblock \href {http://arxiv.org/abs/1007.2581} {\path{arXiv:1007.2581}},
  \href {https://doi.org/10.1016/j.physletb.2010.10.046}
  {\path{doi:10.1016/j.physletb.2010.10.046}}.

\bibitem{Alba:2015iva}
P.~Alba, R.~Bellwied, M.~Bluhm, V.~Mantovani~Sarti, M.~Nahrgang, C.~Ratti,
  {Sensitivity of multiplicity fluctuations to freeze-out conditions in heavy
  ion collisions}, Phys. Rev. C 92~(6) (2015) 064910.
\newblock \href {http://arxiv.org/abs/1504.03262} {\path{arXiv:1504.03262}},
  \href {https://doi.org/10.1103/PhysRevC.92.064910}
  {\path{doi:10.1103/PhysRevC.92.064910}}.

\bibitem{Chatterjee:2016mve}
A.~Chatterjee, S.~Chatterjee, T.~K. Nayak, N.~R. Sahoo, {Diagonal and
  off-diagonal susceptibilities of conserved quantities in relativistic
  heavy-ion collisions}, J. Phys. G 43~(12) (2016) 125103.
\newblock \href {http://arxiv.org/abs/1606.09573} {\path{arXiv:1606.09573}},
  \href {https://doi.org/10.1088/0954-3899/43/12/125103}
  {\path{doi:10.1088/0954-3899/43/12/125103}}.

\bibitem{Bluhm:2018aei}
M.~Bluhm, M.~Nahrgang, {Freeze-out conditions from strangeness observables at
  RHIC}, Eur. Phys. J. C 79~(2) (2019) 155.
\newblock \href {http://arxiv.org/abs/1806.04499} {\path{arXiv:1806.04499}},
  \href {https://doi.org/10.1140/epjc/s10052-019-6661-3}
  {\path{doi:10.1140/epjc/s10052-019-6661-3}}.

\bibitem{Braun-Munzinger:2020jbk}
P.~Braun-Munzinger, B.~Friman, K.~Redlich, A.~Rustamov, J.~Stachel,
  {Relativistic nuclear collisions: Establishing a non-critical baseline for
  fluctuation measurements}, Nucl. Phys. A 1008 (2021) 122141.
\newblock \href {http://arxiv.org/abs/2007.02463} {\path{arXiv:2007.02463}},
  \href {https://doi.org/10.1016/j.nuclphysa.2021.122141}
  {\path{doi:10.1016/j.nuclphysa.2021.122141}}.

\bibitem{Gupta:2022phu}
S.~Gupta, D.~Mallick, D.~K. Mishra, B.~Mohanty, N.~Xu, {Limits of
  thermalization in relativistic heavy ion collisions}, Phys. Lett. B 829
  (2022) 137021.
\newblock \href {https://doi.org/10.1016/j.physletb.2022.137021}
  {\path{doi:10.1016/j.physletb.2022.137021}}.

\bibitem{Asakawa:2019kek}
M.~Asakawa, M.~Kitazawa, B.~M\"uller, {Issues with the search for critical
  point in QCD with relativistic heavy ion collisions}, Phys. Rev. C 101~(3)
  (2020) 034913.
\newblock \href {http://arxiv.org/abs/1912.05840} {\path{arXiv:1912.05840}},
  \href {https://doi.org/10.1103/PhysRevC.101.034913}
  {\path{doi:10.1103/PhysRevC.101.034913}}.

\bibitem{Kitazawa:2013bta}
M.~Kitazawa, M.~Asakawa, H.~Ono, {Non-equilibrium time evolution of higher
  order cumulants of conserved charges and event-by-event analysis}, Phys.
  Lett. B 728 (2014) 386--392.
\newblock \href {http://arxiv.org/abs/1307.2978} {\path{arXiv:1307.2978}},
  \href {https://doi.org/10.1016/j.physletb.2013.12.008}
  {\path{doi:10.1016/j.physletb.2013.12.008}}.

\bibitem{Kitazawa:2015ira}
M.~Kitazawa, {Rapidity window dependences of higher order cumulants and
  diffusion master equation}, Nucl. Phys. A 942 (2015) 65--96.
\newblock \href {http://arxiv.org/abs/1505.04349} {\path{arXiv:1505.04349}},
  \href {https://doi.org/10.1016/j.nuclphysa.2015.07.008}
  {\path{doi:10.1016/j.nuclphysa.2015.07.008}}.

\bibitem{Kitazawa:2012at}
M.~Kitazawa, M.~Asakawa, {Relation between baryon number fluctuations and
  experimentally observed proton number fluctuations in relativistic heavy ion
  collisions}, Phys. Rev. C 86 (2012) 024904, [Erratum: Phys.Rev.C 86, 069902
  (2012)].
\newblock \href {http://arxiv.org/abs/1205.3292} {\path{arXiv:1205.3292}},
  \href {https://doi.org/10.1103/PhysRevC.86.024904}
  {\path{doi:10.1103/PhysRevC.86.024904}}.

\bibitem{Bzdak:2013pha}
A.~Bzdak, V.~Koch, {Local Efficiency Corrections to Higher Order Cumulants},
  Phys. Rev. C 91~(2) (2015) 027901.
\newblock \href {http://arxiv.org/abs/1312.4574} {\path{arXiv:1312.4574}},
  \href {https://doi.org/10.1103/PhysRevC.91.027901}
  {\path{doi:10.1103/PhysRevC.91.027901}}.

\bibitem{Nonaka:2017kko}
T.~Nonaka, M.~Kitazawa, S.~Esumi, {More efficient formulas for efficiency
  correction of cumulants and effect of using averaged efficiency}, Phys. Rev.
  C95~(6) (2017) 064912.
\newblock \href {http://arxiv.org/abs/1702.07106} {\path{arXiv:1702.07106}},
  \href {https://doi.org/10.1103/PhysRevC.95.064912}
  {\path{doi:10.1103/PhysRevC.95.064912}}.

\bibitem{Nonaka:2018mgw}
T.~Nonaka, M.~Kitazawa, S.~Esumi, {A general procedure for
  detector\textendash{}response correction of higher order cumulants}, Nucl.
  Instrum. Meth. A 906 (2018) 10--17.
\newblock \href {http://arxiv.org/abs/1805.00279} {\path{arXiv:1805.00279}},
  \href {https://doi.org/10.1016/j.nima.2018.08.013}
  {\path{doi:10.1016/j.nima.2018.08.013}}.

\bibitem{Gorenstein:2011vq}
M.~I. Gorenstein, M.~Gazdzicki, {Strongly Intensive Quantities}, Phys. Rev. C
  84 (2011) 014904.
\newblock \href {http://arxiv.org/abs/1101.4865} {\path{arXiv:1101.4865}},
  \href {https://doi.org/10.1103/PhysRevC.84.014904}
  {\path{doi:10.1103/PhysRevC.84.014904}}.

\bibitem{Skokov:2012ds}
V.~Skokov, B.~Friman, K.~Redlich, {Volume Fluctuations and Higher Order
  Cumulants of the Net Baryon Number}, Phys. Rev. C 88 (2013) 034911.
\newblock \href {http://arxiv.org/abs/1205.4756} {\path{arXiv:1205.4756}},
  \href {https://doi.org/10.1103/PhysRevC.88.034911}
  {\path{doi:10.1103/PhysRevC.88.034911}}.

\bibitem{Braun-Munzinger:2016yjz}
P.~Braun-Munzinger, A.~Rustamov, J.~Stachel, {Bridging the gap between
  event-by-event fluctuation measurements and theory predictions in
  relativistic nuclear collisions}, Nucl. Phys. A960 (2017) 114--130.
\newblock \href {http://arxiv.org/abs/1612.00702} {\path{arXiv:1612.00702}},
  \href {https://doi.org/10.1016/j.nuclphysa.2017.01.011}
  {\path{doi:10.1016/j.nuclphysa.2017.01.011}}.

\bibitem{Sugiura:2019toh}
T.~Sugiura, T.~Nonaka, S.~Esumi, {Volume fluctuation and multiplicity
  correlation in higher-order cumulants}, Phys. Rev. C100~(4) (2019) 044904.
\newblock \href {http://arxiv.org/abs/1903.02314} {\path{arXiv:1903.02314}},
  \href {https://doi.org/10.1103/PhysRevC.100.044904}
  {\path{doi:10.1103/PhysRevC.100.044904}}.

\bibitem{Sakaida:2014pya}
M.~Sakaida, M.~Asakawa, M.~Kitazawa, {Effects of global charge conservation on
  time evolution of cumulants of conserved charges in relativistic heavy ion
  collisions}, Phys. Rev. C 90~(6) (2014) 064911.
\newblock \href {http://arxiv.org/abs/1409.6866} {\path{arXiv:1409.6866}},
  \href {https://doi.org/10.1103/PhysRevC.90.064911}
  {\path{doi:10.1103/PhysRevC.90.064911}}.

\bibitem{Vovchenko:2020gne}
V.~Vovchenko, R.~V. Poberezhnyuk, V.~Koch, {Cumulants of multiple conserved
  charges and global conservation laws}, JHEP 10 (2020) 089.
\newblock \href {http://arxiv.org/abs/2007.03850} {\path{arXiv:2007.03850}},
  \href {https://doi.org/10.1007/JHEP10(2020)089}
  {\path{doi:10.1007/JHEP10(2020)089}}.

\bibitem{Braun-Munzinger:2019yxj}
P.~Braun-Munzinger, A.~Rustamov, J.~Stachel, {The role of the local
  conservation laws in fluctuations of conserved charges} (7 2019).
\newblock \href {http://arxiv.org/abs/1907.03032} {\path{arXiv:1907.03032}}.

\bibitem{Garg:2017agr}
P.~Garg, D.~Mishra, {Higher moments of net-proton multiplicity distributions in
  a heavy-ion event pile-up scenario}, Phys. Rev. C 96~(4) (2017) 044908.
\newblock \href {http://arxiv.org/abs/1705.01256} {\path{arXiv:1705.01256}},
  \href {https://doi.org/10.1103/PhysRevC.96.044908}
  {\path{doi:10.1103/PhysRevC.96.044908}}.

\bibitem{Sombun:2017bxi}
S.~Sombun, J.~Steinheimer, C.~Herold, A.~Limphirat, Y.~Yan, M.~Bleicher,
  {Higher order net-proton number cumulants dependence on the centrality
  definition and other spurious effects}, J. Phys. G45~(2) (2018) 025101.
\newblock \href {http://arxiv.org/abs/1709.00879} {\path{arXiv:1709.00879}},
  \href {https://doi.org/10.1088/1361-6471/aa9c6c}
  {\path{doi:10.1088/1361-6471/aa9c6c}}.

\bibitem{Nonaka:2020qcm}
T.~Nonaka, et~al., {Pileup corrections on higher-order cumulants}, Nucl.
  Instrum. Meth. A 984 (2020) 164632.
\newblock \href {http://arxiv.org/abs/2006.15809} {\path{arXiv:2006.15809}},
  \href {https://doi.org/10.1016/j.nima.2020.164632}
  {\path{doi:10.1016/j.nima.2020.164632}}.

\bibitem{Kitazawa:2011wh}
M.~Kitazawa, M.~Asakawa, {Revealing baryon number fluctuations from proton
  number fluctuations in relativistic heavy ion collisions}, Phys. Rev. C 85
  (2012) 021901.
\newblock \href {http://arxiv.org/abs/1107.2755} {\path{arXiv:1107.2755}},
  \href {https://doi.org/10.1103/PhysRevC.85.021901}
  {\path{doi:10.1103/PhysRevC.85.021901}}.

\bibitem{Ohnishi:2016bdf}
Y.~Ohnishi, M.~Kitazawa, M.~Asakawa, {Thermal blurring of event-by-event
  fluctuations generated by rapidity conversion}, Phys. Rev. C 94~(4) (2016)
  044905.
\newblock \href {http://arxiv.org/abs/1606.03827} {\path{arXiv:1606.03827}},
  \href {https://doi.org/10.1103/PhysRevC.94.044905}
  {\path{doi:10.1103/PhysRevC.94.044905}}.

\bibitem{STAR:2008med}
B.~I. Abelev, et~al., {Systematic Measurements of Identified Particle Spectra
  in $p p, d^+$ Au and Au+Au Collisions from STAR}, Phys. Rev. C 79 (2009)
  034909.
\newblock \href {http://arxiv.org/abs/0808.2041} {\path{arXiv:0808.2041}},
  \href {https://doi.org/10.1103/PhysRevC.79.034909}
  {\path{doi:10.1103/PhysRevC.79.034909}}.

\bibitem{Alba:2014eba}
P.~Alba, W.~Alberico, R.~Bellwied, M.~Bluhm, V.~Mantovani~Sarti, M.~Nahrgang,
  C.~Ratti, {Freeze-out conditions from net-proton and net-charge fluctuations
  at RHIC}, Phys. Lett. B 738 (2014) 305--310.
\newblock \href {http://arxiv.org/abs/1403.4903} {\path{arXiv:1403.4903}},
  \href {https://doi.org/10.1016/j.physletb.2014.09.052}
  {\path{doi:10.1016/j.physletb.2014.09.052}}.

\bibitem{Vovchenko:2022szk}
V.~Vovchenko, V.~Koch, {Thermodynamic approach to proton number fluctuations in
  baryon-rich heavy-ion matter created at moderate collision energies} (3
  2022).
\newblock \href {http://arxiv.org/abs/2204.00137} {\path{arXiv:2204.00137}}.

\end{thebibliography}

\end{document}